\documentclass[traditabstract]{aa} % for the abstract without structuration                                    % (traditional abstract) 
\usepackage{graphicx}
\usepackage{amsmath}
\usepackage{amssymb}
\usepackage{amsfonts}
\usepackage{textcomp}
\usepackage{txfonts}
\usepackage{natbib}
\usepackage[latin10]{inputenc} %pour linux
\bibpunct{(}{)}{;}{a}{}{,} % to follow the A&A style

\newcommand{\microns}{\ensuremath{\mbox{$\mu$m}}}
\newcommand\normalsubformula[1]{\text{\mathversion{normal}$#1$}}

\begin{document}
%
%   \title{Spectroscopy of faint companions with phase
%   closure nulling} 
   \title{Phase closure nulling} 
   \subtitle{Application to the spectroscopy of faint companions}

   \author{A. Chelli, G. Duvert, F. Malbet \and P. Kern}
   \institute{Laboratoire d'Astrophysique de Grenoble and Mariotti
          Center, UMR 5571 Universit\'e Joseph Fourier, B.P. 53, 38041
          Grenoble Cedex 9, France\\
                        \email{Alain.Chelli@obs.ujf-grenoble.fr}
                    }

   \date{Received; accepted}

   \abstract{We provide a complete theory of the phase closure of
     a binary system in which a small, feeble, and unresolved companion
     acts as a perturbing parameter on the spatial frequency spectrum
     of a dominant, bright, resolved source.  We demonstrate that the
     influence of the companion can be measured with precision by
     measuring the phase closure of the system near the nulls of the
     primary visibility function. In these regions of \emph{phase closure
     nulling}, frequency intervals always exist where the phase closure signature of
     the companion \emph{is larger than any systematic error} and can then
     be measured. We show that this technique allows retrieval of many
     astrophysically relevant properties of faint and close companions
     such as flux, position, and in favorable cases, 
     spectrum. We conclude by a rapid study of the
     potentialities of phase closure nulling observations with current
     interferometers and explore the requirements for a new type of
     dedicated instrument.}
% 5 {} token are mandatory
%   \abstract
%  {}{aims}{method}{result}{} 


   \maketitle
\section{Introduction}

A common feature of the observational problem posed by spectroscopic binaries,
evolved stars companions, hot Jupiters, and extrasolar planets
is the presence of a point-like source of low flux in the vicinity of
a much larger and much brighter star. Indeed, it is the properties of
the faint companion that are desired, and are the most difficult to
acquire. Directly detecting the flux of an extrasolar planet and
establishing its spectrum seems the most demanding task, to be
fulfilled only by dedicated spatial missions
(DARWIN: Cockell et al. 2008, and TPF-I: Lawson et al. 2008)
using interferometry
nulling techniques \citep{1978Natur.274..780B}.

Ground-based optical long baseline interferometers also have the
capability of detecting extrasolar planets restricted to the case of a
giant planet orbiting close to its parent stars
\citep{2006MNRAS.367..825V}, such as 51\,Peg the first exoplanet to be
unraveled by radial velocity techniques \citep{1995Natur.378..355M}.
The principle consists in measuring interferograms in different spectral regions
where the ratio between the planet photons and the stellar ones
changes because of an emission line or an absorption
line in the planet atmosphere. The position of the interferogram
measured by the phase is directly related to the
position of the photocenter of the star-planet system on the sky. The
slight shift in the phase between two spectral regions is in principle
measurable for bright extrasolar planets such as hot Jupiters. 

Even when using color-differential techniques the phase measurement is
plagued by the variation of the refractive index in air. This is why
efforts are currently concentrated on using the phase closure,
which corresponds to the sum of the phases measured in three connected
baselines (see Sect.~\ref{sect:cp} for details). This technique allows
the observer to remove any atmospheric perturbation or
instrumental-based errors localized in each arm of the interferometer.
This detection has not been performed yet because of
instrumental biases and the need to decrease the photon noise even
with large telescopes.

An important property of stars resolved by long baseline
interferometry is that the
coherence of the light decreases with increasing spatial frequencies
down to zero before increasing again following the well known behavior
of the Bessel functions.  \citet{1921ApJ....53..249M} used this
property to measure the diameter of Betelgeuse for the first time.
Indeed, the visibility of a photosphere of uniform brightness
distribution over a circular disk is canceled out at the spatial
frequency $0.61/R_\star$ where $R_\star$ is the stellar angular radius.
Another property is that the phase of the visibility jumps by 180
degrees at each crossing of the Bessel function's zeros (hereafter
called visibility nulls or simply nulls). At each null, the coherent flux of
the star is canceled out, making it easier to detect a change in the phase 
jump due to a faint close companion.

In this paper we demonstrate that, in precisely the restricted case of
the lobe crossing, the influence of even a very faint companion can be
measured with precision, and its spectrum can be obtained by measuring the phase
closure of the system near the nulls of the primary visibility
function. In these regions where the \emph{phase closure nulling} of
the primary is effective, the phase closure signature of the secondary
stands out to a point where its basic properties (flux, position,
spectrum, etc.) can be measured.  In a companion paper \citep{paperII},
this technique was used to detect the companion of the single-lined
spectroscopic binary $\sigma$\,Puppis, at a few stellar
radii distance.
\section{Description of the problem}
\begin{figure*}[t]
    \centering
\includegraphics[width=0.3\hsize,angle=-270]{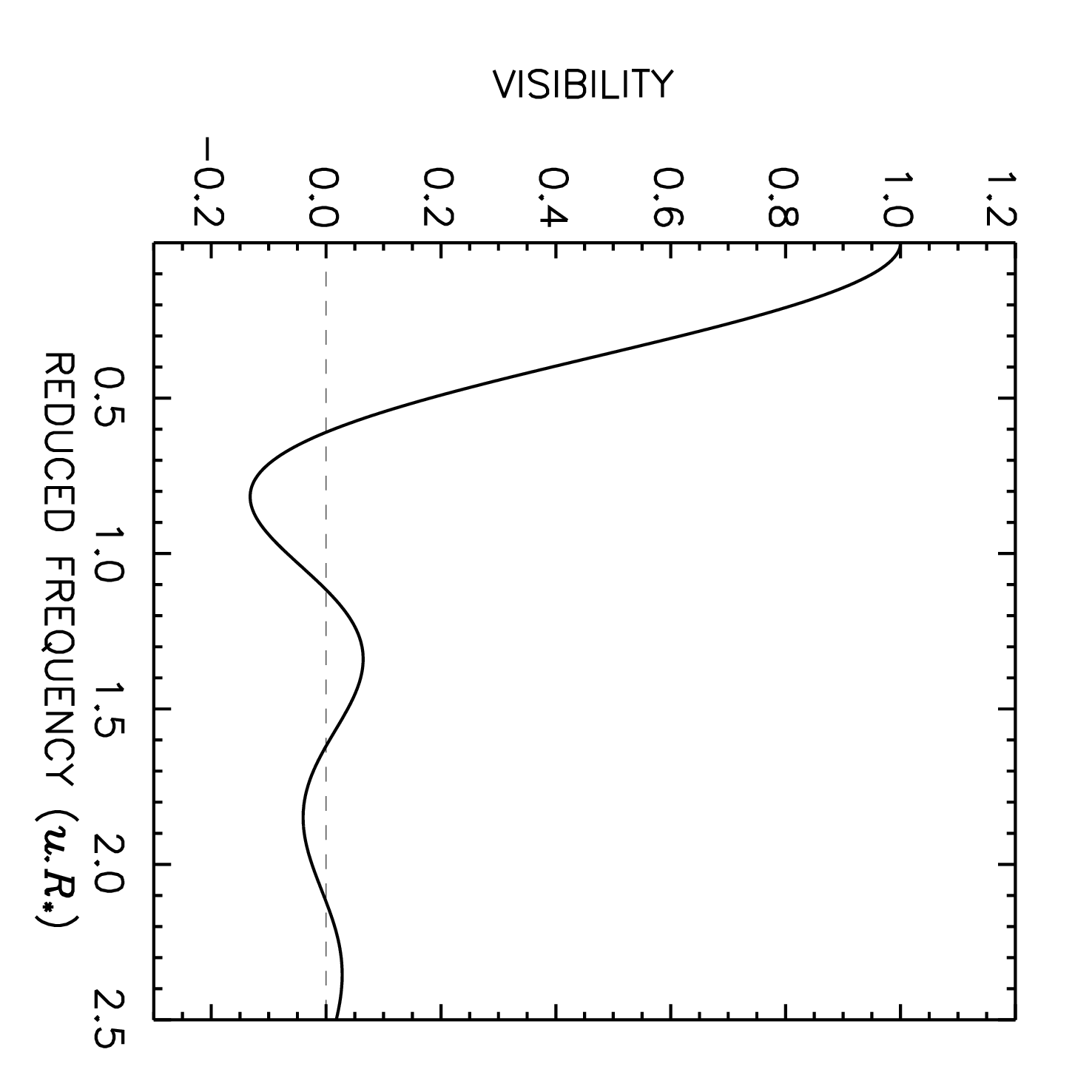} \hfill
\includegraphics[width=0.3\hsize,angle=-270]{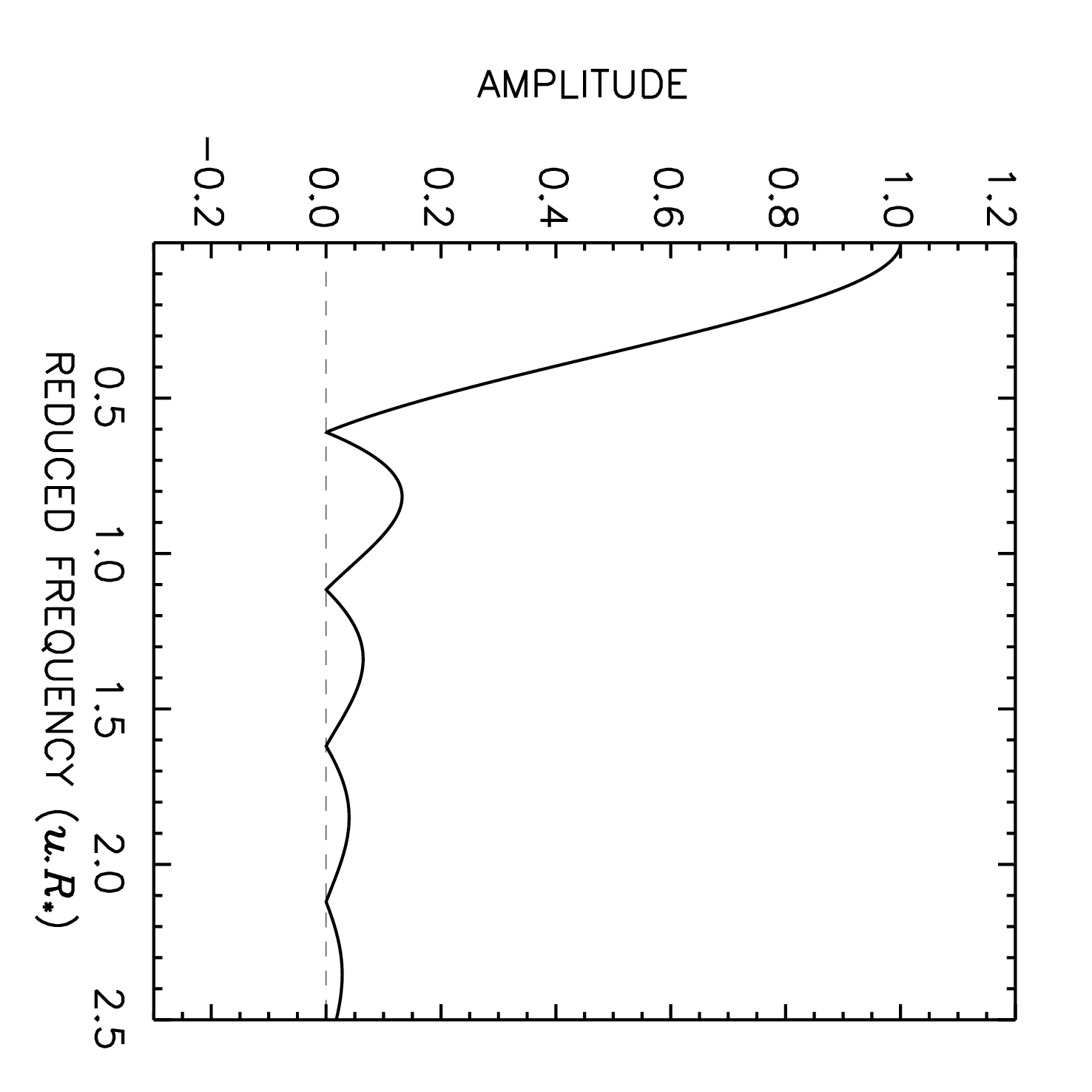} \hfill
\includegraphics[width=0.3\hsize,angle=-270]{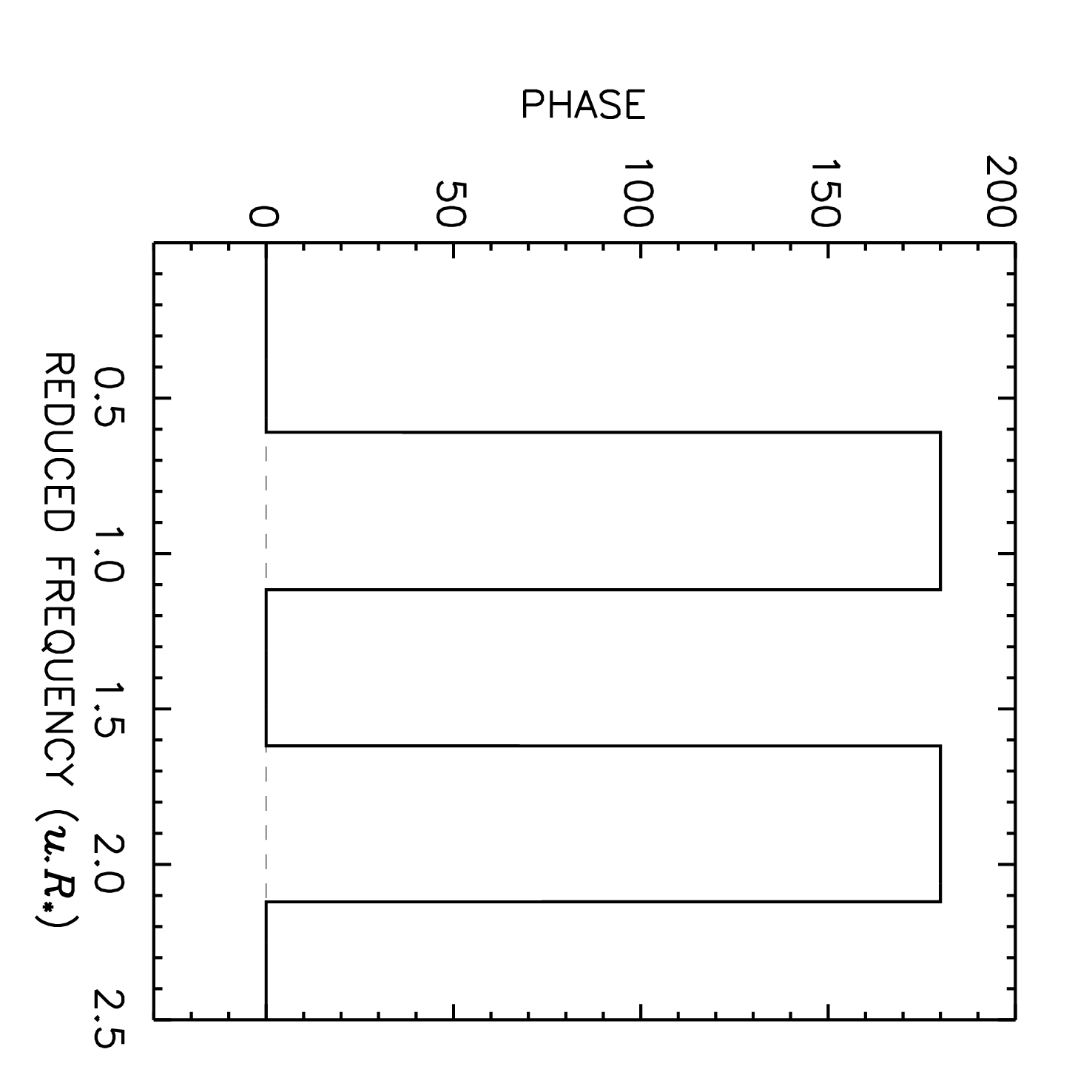}\\
\caption{Visibility for a star well represented by a uniform
  brightness distribution over a disk of angular radius $R_\star$(left
  panel), as a function of the reduced frequency (defined as the
  frequency times the stellar radius). The visibility is often
  presented under the form of 
  amplitude (center panel) and phase (right panel).}
\label{fig:bessel}
\end{figure*}
\subsection{Visibility of a stellar photosphere of uniform brightness}
\label{sec:UD_star}
When observing a star with a multi-aperture
interferometer, it is well known that one detects the amplitude of the
visibility function (or the amplitude squared) and that this amplitude
decreases with increasing spatial frequency $u=B/\lambda$, where $B$
is the baseline and $\lambda$ the wavelength. Thanks to the
Zernike-Van Cittert theorem, it can be shown that this visibility
function coincides with the Fourier transform of the spatial
energy distribution of the star; hence, for a star whose energy distribution
is well-represented by a uniform disk, the visibility function
is proportional to a Bessel function of the first order:
\begin{equation}
  \label{eq:UD}
  V_\star(u) = 2\,\frac{J_1 (2\pi\,u\,R_\star)}{2\pi\,u\,R_\star}.
\end{equation}
The visibility function is a complex number that, in this special case
of a simple centro-symmetric brightness distribution, turns out to be
real (i.e.\ the imaginary part is zero). It is
often displayed by its amplitude and its phase (see
Fig.~\ref{fig:bessel}).

The visibility becomes negative when the baseline is longer
than $0.61\,\lambda / R_\star$, corresponding to the first null, so
the phase suddenly changes from zero to  
180 degrees (see right panel of Fig.~\ref{fig:bessel}). It becomes
positive again when crossing the second null, and so on.

\subsection{Visibility of a resolved star  with a faint companion}
\label{sec:faint_binary}
When an object is composed of two components, the object visibility
function is the sum of the visibility function of the two components
weighted by the fraction of flux of each component and a phase
modulated by the apparent angular separation on the sky. If one
adds a faint companion to the star of the previous section with a
flux ratio  $r$ and separated from the star by $s$
(assuming for sake of simplicity that the system orientation is
parallel to the frequency axis), then the complex visibility of the
system is
\begin{equation}
  \label{eq:binary}
  \hat{\i}(u)=\frac{V_\star(u)+r e^{i\,2\pi\,us}}{1+r}.
%  V(B) = (1-f)\,V_s(B) + f\,\exp(2\pi\,j\,sB/\lambda) 
\end{equation}
If the flux ratio $r$ is small, the amplitude of the visibility
function will be very slightly modified by the presence of the faint
companion. The stronger effect occurs around visibility nulls of the
primary, where the system visibility is no longer zero. At these
locations, the visibility perturbation is equal to the flux
ratio. This effect still remains weak and is beyond the
performances of current interferometers as soon as the flux ratio is
less than about 1\%. However, besides the fact the companion is
not detectable, studies around visibility nulls keep their interest for
precise diameter and limb darkening measurements \citep{2008A&A...481..553L}. 
\subsection{Phase changed by the faint companion}

More interesting is the phase of the object visibility.
By definition, the tangent of the
phase of a complex number is equal to the ratio of its imaginary part
by its real part. Therefore we have
\begin{equation}  
\tan\phi(u) = \frac{r\,\sin(2\pi\,us)}{V_\star(u) +
    r\,\cos(2\pi\,us)}.
\label{eq:phase}
\end{equation}
For small flux ratios and except for around nulls, the phase of
the object is very close to 0 or 180 degrees, and may be approximated
by
\begin{equation}
\label{eq:phase_approx}
\phi(u) \approx \frac{r \sin(2\pi\,us)}{V_\star(u)}+n\pi
\end{equation}
with $n=0$ if $V_\star(u){>}0$, and $n=1$ if $V_\star(u){<}0$. For
separations $s/R_\star$ much greater than 1, it is quasi periodic, with a
pseudo period $s^{-1}$.  

The phase is the result of two contributions: from the star
$V_\star(u)$ and from the companion $r\,e^{i\,2\pi\,us}$. The
contribution of the star decreases as the
frequency increases and is even canceled at the nulls, while that of the
companion keeps constant as long as it is not resolved. Most of the
time, the companion produces a phase signature in the range $\pm
r/V_\star(u)$. But more important,  
around visibility nulls, in the frequency ranges for which $|V_\star(u)|
\le r$, this phase signature becomes significant, with an exact 
value of $2\pi\,u_0 s$ at the frequencies $u_0$ of the nulls (${\approx}
\pi s/R_\star$ for the 
first null), much greater than 180 degrees. It follows
that, as opposed to the visibility amplitude, even for small flux
ratios, frequency intervals always exist around nulls, where the 
phase signature of the companion is larger than any systematic error
and can then be measured.

\subsection{Phase closure of a resolved star with a faint companion}
\label{sect:cp}
An interferometer formed by 3 telescopes transmits 3 frequency
systems, $u_{12}$, $u_{23}$, and $u_{13}=u_{12}+u_{23}$, associated
with the 3 baselines, each characterized by a phase and a
visibility. It provides one phase closure defined as the phase of the bispectrum
$\hat{I}(u_{12},u_{23})$, with
\begin{equation}
\hat{I}(u_{12},u_{23})=<\hat{\i}(u_{12})\hat{\i}(u_{23}){\hat{\i}}^*(u_{13})>
\label{bispectrum}
\end{equation}     
where ${}^*$ denotes the complex
conjugate, and $<>$ represents an ensemble
average. The phase closure has two important 
properties: it is insensitive to atmospheric phase fluctuations, and
for a point source, it is equal to zero. Then, for an extended object,
it is given by
\begin{equation}
\phi_c=\phi_o(u_{12})+\phi_o(u_{23})-\phi_o(u_{13})
\end{equation}     
where $\phi_o$ is the phase of the object spatial Fourier
transform. 

If the telescopes are aligned with baselines parallel to the direction
of the double system, the phase closure may be written
\begin{eqnarray}
\phi_c \approx \cfrac{rsin(2\pi u_{12}s)}{V_\star(u_{12})}+ 
 \cfrac{rsin(2\pi u_{23}s)}{V_\star(u_{23})} - \cfrac{rsin(2\pi
 u_{13}s)}{V_\star(u_{13})}+\nonumber\\
(n_{12}+n_{23}-n_{13})\pi. 
\label{clot}
\end{eqnarray}
\begin{figure}
  \includegraphics[width=0.55\columnwidth,angle=-270]{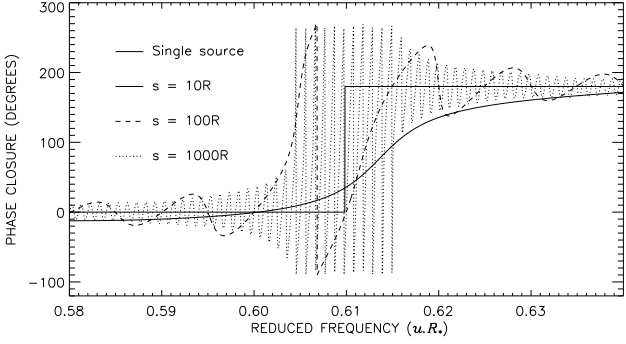}
  \caption{Phase closure modulo $2\pi$ of a double system, formed by an extended
    uniform disk and a point source, as a function of the
    maximum reduced frequency. The spectral resolution is
    $\mathcal{R}=1500$. The 3 frequencies \{$u_{23}, u_{12}, u_{13}$\}
    (see Eq.~\ref{bispectrum})
    have been chosen in the
    ratio 1,2,3, with the maximum frequency around the first zero of
    visibility of the primary star. The phase closure is displayed
    for 3 separations ($10\,R_\star$, $100\,R_\star$, and $1000\,R_\star$) and
    a flux ratio of $0.01$. The signature of the secondary
    source is dominant in the region around the minimum visibility
    of the primary (near reduced frequency $0.61$). The thin line with a 180 degree shift
    at 0.61 corresponds to the phase closure of the primary alone.} 
  \label{fig:closure_phase}
\end{figure}
Figure \ref{fig:closure_phase} shows the expected phase closure, around
the first visibility null, from a double system with a flux ratio of
1\% and various separations. The system is observed with  a spectral 
resolution of 1500, like the one available with the AMBER instrument
at the VLTI \citep{2007A&A...464....1P}. Note the importance of the
phase closure signature from the companion. 
\section{Phase closure nulling}
\label{sect:phaseclosure}
In the case of a system formed by an extended uniform disk and
a point source, the useful phase closure signal is the departure from 0 or 180
degrees. We have then clearly interest in selecting the working frequency interval
at the highest frequencies, around visibility nulls at $u=u_0$, where
the phase closure signature of the companion is at its maximum. In these
regions, there is always (as 
the stellar visibility crosses the zero) a frequency interval within
which the useful signal is stronger than any systematic error, hence
can be measured. 

\begin{figure}
  \includegraphics[width=\columnwidth,angle=-270]{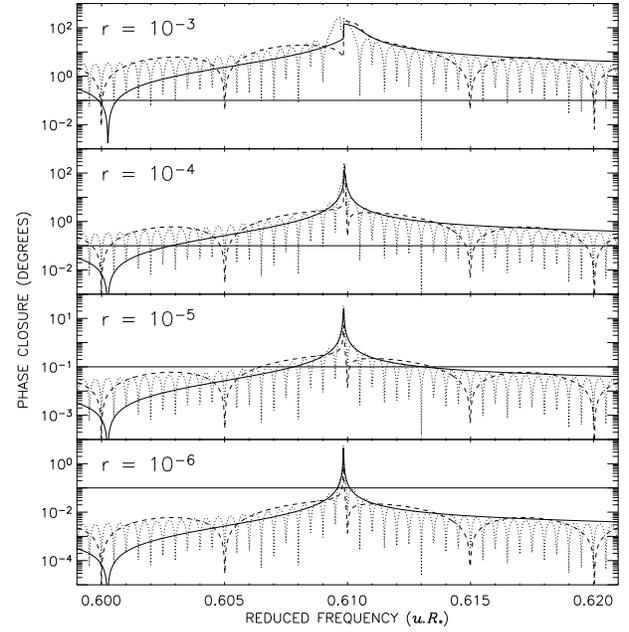}
  \caption{Absolute value of the useful phase closure signal of a double system
    formed by a uniform disk and a point source as a function of the
    highest frequency located around the first visibility null of the
    primary. The spectral resolution is $\mathcal{R}=1500$. The
    phase closure of the uniform disk has been subtracted for clarity.
    From top to bottom, the flux ratio varies from $10^{-3}$ to
    $10^{-6}$, and 3 separations are considered: $10\,R_\star$ (full
    lines), $100\,R_\star$ (dashed lines), $1000\,R_\star$ (dotted lines).
    The horizontal lines are set to a given systematic error of
    0.1 degree, showing the regions where the phase closure signature
    of the secondary becomes measurable.}
  \label{fig:closure_phase_2}
\end{figure}

Figure~\ref{fig:closure_phase_2} shows the useful phase closure signal
for various flux ratios and distances of 10, 100, and 1000 stellar
radii. The horizontal lines correspond to a given systematic error of
0.1 degree. The frequency interval within which 
the useful signal is stronger than this error diminishes rapidly with
the flux ratio, but is always present.

We propose using the regions around minima of visibility, where the
{\it phase closure nulling} of the primary is effective, and the
useful signal is over the limit $\sigma_{sys}$ imposed by
systematic errors, to detect and to characterize faint
companions. By doing so, we do not cancel the stellar flux
as in classical nulling experiments \citep{1978Natur.274..780B}, but
we only cancel the stellar coherent flux. The main limiting noise will then be
the photon noise from the central star, as shown in Sect.~\ref{sect:errors}. 
\subsection{Spectroscopy of close companions}
The spectroscopy of close companions may be extracted from spectrally dispersed
phase closure measurements obtained with an $N_{{\normalsubformula{\text{tel}}}}$ telescopes interferometer
($N_{{\normalsubformula{\text{tel}}}}{\ge}3$). The baselines should be such that the longest one
corresponds to a null of the primary star in order to maximize the
useful signal. As an example, an interferometer with 3
telescopes and frequencies \{$u_{23}, u_{12}, u_{13}$\} in the ratio
1:2:3, allowing bootstrapping, 
would be a very efficient system. The data consists of spectrally
dispersed phase closure measurements performed as a function of the
spatial frequency.    

The double system is characterized by 4 parameters: the stellar radius
$R_\star$, the flux ratio $r$, the separation $s$, and the position
angle. The 4 parameters may be extracted from a  
modeling of the phase closure variations as a function of both the spatial
frequency and the wavelength. The spectrum of the companion is then
obtained by multiplying the
flux ratio by the spectrum of the primary. 
\subsection{Error analysis}
\label{sect:errors}
We consider for simplicity an interferometer formed by $N_{{\normalsubformula{\text{tel}}}}$
aligned telescopes in a non-redundant configuration. We also assume that
the direction of observation varies little during the
experiment so that the double system is characterized by only 3
parameters:  the stellar angular radius $R_\star$, the flux ratio $r$, and the
projected separation $s$ along the direction of
observation. Spectrally dispersed phase closure   
measurements are performed as a function of the spatial frequency and
then fitted, with a 3-parameter model. 

The limiting noises in optical interferometry from the visible to the
near infrared are the atmospheric noise, the photon noise from the 
source, and the detector readout noise. General
expressions of the errors on the parameters of the double system are derived in
Appendix~\ref{ann:errors}. In  the photon noise regime, the errors reach 
their minimum values and then stay constant as soon as the working
frequency domains cover at least one period $s^{-1}$. For separations greater than a few
stellar radii and working frequency intervals over $s^{-1}$, the error on the
flux ratio and that on the separation, with a 3-telescope
interferometer, are given by
\begin{eqnarray}
\sigma(r) \approx & \cfrac{3}{\sqrt{K}} &\\
\sigma(s/R_\star) \approx & \cfrac{1}{r\,\sqrt{K}}~ &
\biggl(\cfrac{0.61}{\bar{u}_{max}\,R_\star}\biggr)    
\end{eqnarray}
where $K$ is the total number of photoevents detected during the whole
observation and $\bar{u}_{max}$ the average frequency of the
longest baseline.

These formulae have been checked numerically through a rigorous error
analysis. They are valid everywhere, even around the nulls. Besides
the error    
on the separation being inversely proportional to the maximum frequency,
the errors are independent of the working frequency domains. This
clearly reinforces the usefulness of observing around
visibility nulls, where the useful phase closure signal is dominant. The
error on the flux ratio only depends on the total number of collected
photoevents, while that on the distance is also inversely proportional
to the flux ratio. In the photon noise regime, both errors are
independent of the spectral 
resolution $\mathcal{R}$, as long as the spectral averaging is
performed over less than half a period $1/2\,s$, that is, when
$\bar{u}_{max}\mathcal{R}^{-1}{<} 0.5\,s^{-1}$. This condition provides a
maximum recoverable distance imposed by the spectral resolution of
\begin{equation}
\frac{s_{max}}{R_\star} \approx
\mathcal{R}\times\biggl(\frac{0.61}{\bar{u}_{max}R_\star} \biggr ) \, .
\end{equation}
The error on the flux of the companion is
\begin{equation}
\sigma(rK) \approx 3\sqrt{K} \, .
\end{equation}
The direct photometric detection of the companion would provide an
error of $\sqrt{rK}$. In terms of performances, the direct detection is
$3/\sqrt{r}$ better than detection from phase
closure. However, 
this is the price to pay because, to our knowledge, there is no other
direct method than the one we propose, which is capable of detecting stellar
companions within an Airy disk and at a distance from the hosting star
as small as a few stellar radii. Once the conditions above are
fulfilled, measuring the flux ratio for each spectral element at
resolution $\mathcal{R}$ is an effective way to measure the
companion spectrum.
\begin{figure}
  \includegraphics[width=1.1\columnwidth,angle=-270]{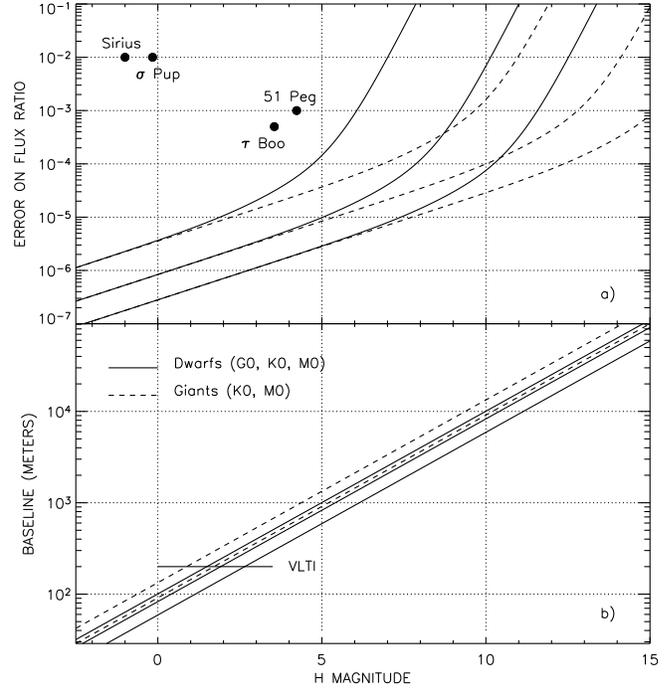}
  \caption{~\textbf{a} (top): Error on the flux ratio of a double system from the closure
    phase information of a 3-telescope interferometer. The signal is made by $900$ phase
    closure regularly sampled within the working frequency intervals,
    and spectrally dispersed along $270$ spectral channels, with a
    spectral resolution of $1500$. The experimental parameters are;
    observing wavelength: $1.65 \microns$; optical
      bandwidth: $0.3 \microns$; total integration time: 3 hours;
    integration time per frame: $0.2$s; detector
      readout noise: $10e^-$; $32$ pixels per frame. The calculations
    took the photon noise and the
    detector readout noise into account. Three cases were considered: 3
    telescopes with diameter D=$2$m, a Strehl ratio S=$0.5$, and $1$\% total
    transmission (upper curve, this case corresponds roughly to the
    present state of the AMBER experiment on the VLTI), D=$2$m,
    S=$0.9$ and $10$\% total transmission (middle curve), D=$8$m,
    S=$0.5$ and $10$\% transmission (lower curve). The broken lines
    correspond to an integration time per frame of $12$ seconds.
    Represented are: Sirius with a possible M dwarf companion \citep{1995A&A...299..621B}, the giant K5
    star $\sigma$ Puppis with its G2 dwarf companion \citep{paperII}, and two stars
    with known planetary companions.  
    \textbf{b} (bottom): Baseline needed to resolve
     the primary star in the H band at $1.65\mu m$, namely to reach
    the first visibility null, for late dwarfs and giants. The
    horizontal line at $200$\,m corresponds to the maximum baseline of the
    VLTI.}  
  \label{fig:flux_ratio}
\end{figure}
\subsection{Performances}
The mean error on the flux ratio in the H band at $1.65 \microns$
from a 3-telescope interferometer and an integration time of 3 
hours, obtained 
by averaging all the spectral channels over an optical bandwidth of
$0.3\microns$, is plotted in Fig.~\ref{fig:flux_ratio}a. The upper
curves correspond to 3 telescopes of $2$\,m, 1\% transmission, and a
Strehl ratio of 0.5. It roughly represents the present state of the
AMBER instrument on the VLTI with Auxiliary telescopes. The middle and
lower curves correspond
to telescopes of $2$\,m and $8$\,m, a transmission of 10\%, and Strehl
ratios of 0.9 and 0.5, respectively (see the caption of
Fig.~\ref{fig:flux_ratio} for details).   

If we assume a signal-to-noise ratio of 10 on the flux ratio for a
positive detection, then the AMBER instrument with Auxiliary
telescopes would allow detection of companions $10^4$ fainter around
second-magnitude stars and $10^3$ fainter around five-magnitude
stars. Spectroscopy would be possible from flux ratios of
$10^{-4}$. Optimized interferometers with 10\% 
transmission (Fig.~\ref{fig:flux_ratio}, middle and bottom curves)
would allow the detection of companions $10^4$ fainter around
5 to 7 magnitude stars and $10^3$ fainter around 8 to 10 magnitude 
stars. Spectroscopy would be possible from flux ratios of $10^{-5}$.    

Figure~\ref{fig:flux_ratio}b shows the baseline needed to resolve
the primary star, that is, to reach the first visibility null for late
dwarfs and giants. The horizontal line at $200$\,m corresponds to the
maximum baseline of the VLTI, which imposes maximum
H magnitudes of about $2.5$ and $1.5$ for M0 dwarves and giants,
respectively. Also the field of view of fiber-linked interferometers is limited to one Airy
disk $\lambda/D$, where $D$ is the telescope diameter. Given that the
baseline for reaching the first null is $B=0.61\lambda/R_\star$, it gives
a maximum recoverable separation of about $B/D$ stellar radii,
which is 100 stellar radii on the VLTI with $2$\,m telescopes. A baseline
of $1000$\,m necessary for resolving a solar type star at a distance of 10pc
would provide a field of 500 stellar radii. 

As specified in Sect.~\ref{sect:errors}, the minimum error is attained
in the photon noise regime when the
working frequency intervals cover at least one period $s^{-1}$. This
implies, for the first null, a baseline variation $\Delta B$ of
\begin{equation}
\Delta B \approx 2\frac {B}{s/R_\star}.
\end{equation}
At kilometric baselines, it represents a variation of $200$\,m for a
distance of 10 stellar radii. The simplest way to achieve such
variations is to use earth rotation with repositionable
telescopes located at positions such that the baselines and
orientations optimize the observing time around the nulls. However,
the double system may also be characterized at the price of a lower
signal-to-noise ratio, even if the working frequency intervals cover
much less than a period. Indeed, the companion of the spectroscopic
binary $\sigma$ Puppis has been detected and characterized with only 3
snapshot observations of 5 minutes each; before, during and after the
null, using the spectral dimension to increase the frequency
coverage. 
\section{Discussion}
\begin{figure}
  \includegraphics[width=0.65\columnwidth,angle=-90]{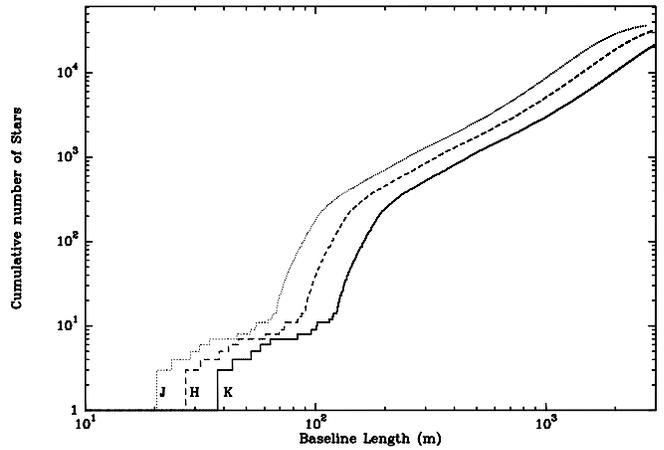}
  \caption{Lower limit of the number of stars resolved at a given baseline in
    the J, H, and K bands.}
  \label{fig:cumulative}
\end{figure}

\subsection{Science cases}
We have demonstrated that a phase closure nulling experiment can bring
astrophysically significant information on the close environment of
resolved stars.  Figure~\ref{fig:cumulative} shows the lower limit
of the number of targets available to phase closure nulling
experiments as a function of baseline length.  The sample was obtained with the SearchCal utility
    \citep{2006A&A...456..789B} and is limited to the stars whose
    apparent diameter can be reliably estimated from parallax and
    (spectro)photometric measurements available at the CDS.
The minimum number of targets available for current arrays like the VLTI is
${\approx}100$ in K and ${\approx}500$ in J.  The technique is not
limited to observing unbalanced binary systems but can also
address a broader range of topics.  These topics may change either by
changing the nature of the resolved bright astrophysical target and/or
the off-axis accompanying astrophysical signal.

In star and planet formation, astronomers are interested in
protoplanetary disks, which in general have the property of being
centro-symmetric so that the phase closure signal is zero. Any
perturbating signal located off-axis can then be detected by closure
phase nulling: inhomogeneities on the disk surface, presence of a
forming planetesimal, presence of a collimated jet of outflowing wind
partially screened by the disk. It also applies to other science cases
holding disk-like cataclysmic variables, active galactic nuclei,
proplanetary nebulae, gaseous disk around hot stars.  The case of
exoplanets is of particular interest, but the various kind of
binaries must not be forgotten involving young, main sequence, or evolved
stars and even the case of merging binaries.

In a future where interferometry is sensitive to fainter targets,
a possible application of this technique is the detection of the
gravitational images due to a massive gravitational lens spatially
resolved. In the following sections, we restrict ourselves to the
companion of evolved and main sequence stars. 
\subsubsection{Companions of evolved stars}
Clearly the simplest observational case, already permitted with
current interferometers, is the study of \emph{close companions to
  giant stars}. The scientific interest is twofold: improving our
knowledge of the masses of giant stars and detecting planetary
systems around earlier type stars than the ones known to date.  It is
indeed customary to look for companions of stars of
masses~${>}1.2\mathrm{R}_{\sun}$, not in their main-sequence phase,
where radial velocity measurements are rendered difficult by the
paucity of spectral lines (furthermore broadened by the rapid stellar
rotation), but in their giant, cooler phase. However, the detection of
substellar companions around giant stars using radial velocities has
always been problematic because the intrinsic pulsation of those stars
would mimic the signature of a close substellar companion, and radial
profiles can be altered by large, slowly rotating,
cool spots on the surface.  Only a few detections have been reported
to date \citep{1998MNRAS.293..469H,2005A&A...437..743H,
  2007A&A...472..649D}. The phase closure nulling would provide a
completely independent method of characterizing at least the closest
companions, and give the orbit needed to measure individual masses.

We demonstrate this very possibility in a companion paper
\citep{paperII}, where we apply the
phase closure nulling method for the detection of the close companion
of the star Sigma\,Puppis and the 
characterization of this SB1 binary system.
\subsubsection{Companions of main-sequence stars}
These companions can range from brown dwarves (hereafter BDs) or hot
Jupiters (HJs) to
earth-like planets. With an expected flux ratio of a few $10^{-4}$ in
the near infrared, brown dwarves could be
detected with present-day telescope apertures at any distance from the
star within the field-of-view permitted (${\approx}B/D$). The
experiment could bring two new pieces of information: the mass of the BDs (by
solving the orbit) and the NIR spectrum, possibly different in close
binaries than the spectra of isolated BDs. 

A few HJs, thus very close to their star (${<}20\mathrm{R}_\star$) and
with temperatures of $1000$--$1500$\,K, will be detectable as well
with a flux ratio in the $10^{-4}$--$10^{-3}$ range in K band and
longwards \citep[see e.g.][]{2008ApJ...678.1419F}.  A handful of HJs
have already been photometrically detected with the IRAC and MIPS
instrument of the Spitzer Space Telescope
\citep{2005ApJ...626..523C,2008ApJ...673..526K} in transit
experiments.  Phase closure nulling experiments in this case would
permit the NIR spectrum of the object, the orbit and thus
the mass to be obtained, without being restricted to transit objects. Phase closure
nulling experiments on dwarf stars with HJs would only be possible 
today on a handful of them since 
the baseline needed to null a Sun equivalent at $10$\,pc is
${\approx}1000$\,m at $1.65\microns$. This restriction can be overcome in the near
future as fibered interferometers with kilometer baselines are
currently underway \citep{2006Sci...311..194P}. 

For any planetary companions still cooler and smaller than HJs, the flux
ratio between the star and the companion rapidly becomes forbidding
(${>}10^9$) in the NIR and is manageable only longwards of
$10\,\microns$ \citep[see Fig.~2 in ][]{2002assc.book..369T}.  In the
N band, the above-mentioned constraint on the baseline length then becomes very
challenging (${>}10^4$\,m), and the
signal transmission on such distances would be quite impossible for
ground-based interferometers, 
unless techniques like the parametric conversion to visible or NIR
wavelength, such as described in \citet{2008OptCo.281.2722D}, are used
before recombination.

\subsection{Towards a dedicated instrument}
We have discussed the main properties
of phase closure nulling: around each null of the visibility function
of the star, a spatial frequency interval exists where the
signal of a companion becomes stronger than any systematic error, hence
potentially detectable. The errors on the companion detection, both in
flux ratio and in position, depend roughly on the square root of the
number of star photons detected, and the imageable region for a
fiber-linked interferometer (the
distance from the star where the companion's signature can be seen) is
${\approx}B/D$ stellar radii. The need to be at a star null fixes
a minimum baseline length. The need to reach a given sensitivity for
the flux ratio implies either a given minimum size for the telescope
apertures, thus a maximum imageable area, or/and a given integration time. 
The spectral resolution $\mathcal{R}$ does not need to be very
high, typically ${\approx}B/D$. 

Today interferometers with baselines of a few hundred meters can
already detect faint companions to the few closest giant stars using
phase closure nulling. It seems possible to go further --- and still
use ground-based optical arrays --- to characterize the faint
companions, down to HJ masses, of the nearby main-sequence stars.
This can be envisioned with an interferometer, that
\begin{enumerate}
\item reaches the first zero of a solar-type
  star at $10$\,pc in an infrared band --- already kilometric baselines in H band;
\item has a spectral resolution of ${\approx}B/D$ to enable detection of
  companions up to ${\approx}B/D$ stellar radii. This is $500$ for a $1000$m
  baseline and $2$m-class telescopes;
\item permits long integration times on the zero crossing thanks
  to earth rotation and repositionable telescopes;
\item has the highest attainable transmission: since our S/N
  only depends on the photon S/N of the primary, this should be as
  large as possible. Since we are only interested in phase closure
  that is a robust measurement insensitive to relay optics wavefront
  disturbance or atmosphere stability, the attention should go on
  improving the transmission of the relay optics, not their optical
  quality. This is best realized with fully integrated optics,
  fiber linking the foci of the telescopes at the primary focus.
\item uses a few detectors with low noise for the detection, such
  as a 4-pixel ABCD method. New developments
  \citep{2007SPIE.6542E..45R} seem to imply that such 
  low-noise IR detectors can be available in a few years from now.
\item takes advantage of the new developments in integrated
  optics, permitting both beam recombination and spectrography, thus
  avoiding the losses at all the diopters of bulk optics used for beam
  transportation and in the spectrograph. Such a concept,
derived from on-going developments for visible spectroscopy
\citep{LeCoarer07} is under investigation \citep{Kern09}
and can be envision on the sky in the coming few years.
\end{enumerate}

\section{Conclusion}
We have introduced both the concept of phase closure nulling and its
application in the detection of faint close companions to resolved
stars, accuratly measuring their position and spectra. This 
concept is the only one to date to permit such direct detection
\emph{inside} an Airy disk. We developed the complete error
analysis of the phase closure nulling, and show how this method
permits ground-based interferometry to complement and, in favorable
cases, compete with the direct detection methods considered for future
space-borne experiments. We discussed a few science cases that
would 
benefit from phase closure nulling experiments and sketched the
requirements for a new type of interferometer dedicated to such
studies.
\appendix
\section{Error calculation}
\label{ann:errors}
\subsection{Error on phase closure}
The formal error on the phase closure, in the presence of photon
noise and detector readout noise, was computed by \citet{1989A&A...225..277C} and
\citet{2005JOSAA..22.1589T}. Neglecting the coupling between the 
photon noise and the detector noise and retaining the dominant terms,
the variance of the phase closure $\phi_c$ may be approximated by
\begin{eqnarray}
\sigma^2(\phi_c) & \approx &
\frac{N_{{\normalsubformula{\text{tel}}}}^2}{2K} \times \biggl 
[\frac{1}{|V(u_{12})|^2}+\frac{1}{|V(
    u_{23})|^2}+\frac{1}{|V(u_{13})|^2}
\biggr ] \, \nonumber \\
& + & \frac{1}{2}\biggr (
\frac{N_{\normalsubformula{\text{tel}}}^2N_{pix}\sigma_r^2}{K^2}
\biggl )^3  \times
\frac{1}{|V(u_{12})|^2|V(u_{23})|^2 |V(u_{13})|^2}  \,    ,
\end{eqnarray}
where $K$ is the total number of detected photoevents,
$N_{{\normalsubformula{\text{tel}}}}$ the number of  
telescopes, $N_{pix}$ the number of pixels, and
$\sigma_r$ the detector readout noise. And $|V(u_{12})|$,  $|V(u_{23})|$,
$|V(u_{13})|$, are the visibility amplitudes of the observed object. 

\subsection{Error on the parameters of a double system}
We want to estimate the errors on the parameters of a double system
made of an extended uniform disk and a point source, 
from a least square fit of a set of $N_c$ uncorrelated phase closure 
measurements obtained with an interferometer formed by $N_{{\normalsubformula{\text{tel}}}}$
telescopes. For simplicity, we assume that the telescopes are aligned,
with a non redundant configuration, and that the direction of
observation varies little during the experiment and may be considered
constant. Under these conditions, the object is characterized by 3
parameters: the stellar angular radius $R_\star$, the flux ratio $r$, and the
projected separation $s$ along the direction of observation. 

The $N_c$ phase closures are sampled at frequencies  
$u_{ln}^k \{k=1,...,N_c; ln=12,23,13\}$, in frequency intervals
$\Delta u_{ln}$. We define the $N_c$ lines and 3 columns matrix
$\mathcal{M}$, of element $m_{jk}$, by
\begin{equation}
m_{jk}=\frac{1}{\sigma(\phi_c)}\cfrac{\partial\phi_c^k}{\partial
  p_j}~~~~({\rm with}~\{p_1,p_2,p_3\}=\{R_\star,r,s\}).
\end{equation}
The variances of the parameters are the diagonal elements
of the matrix $[^t\mathcal{M}\times\mathcal{M}]^{-1}$. Let us first
examamine the matrix  $^t\mathcal{M}\times\mathcal{M}$. It is given by
\begin{equation}
\begin{bmatrix}
\sum_{k} {\cfrac{1}{\mathit{{\sigma}}_k^2}\Bigl(\cfrac{\partial\phi_c^k}{\partial R_\star}\Bigr)^2}&
\sum_k {\cfrac{1}{\mathit{{\sigma}}_k^2}\cfrac{\partial\phi_c^k}{\partial R_\star}\cfrac{\partial\phi_c^k}{\partial{r}}}&
\sum_k {\cfrac{1}{\mathit{{\sigma}}_k^2}\cfrac{\partial\phi_c^k}{\partial R_\star}\cfrac{\partial\phi_c^k}{\partial{s}}}\\
\sum_k {\cfrac{1}{\mathit{{\sigma}}_k^2}\cfrac{\partial\phi_c^k}{\partial R_\star}\cfrac{\partial\phi_c^k}{\partial{r}}}&
\sum_k {\cfrac{1}{\mathit{{\sigma}}_k^2}\Bigl(\cfrac{\partial\phi_c^k}{\partial r}\Bigr)^2}&                  
\sum_k {\cfrac{1}{\mathit{{\sigma}}_k^2}\cfrac{\partial\phi_c^k}{\partial s}\cfrac{\partial\phi_c^k}{\partial{r}}}\\
\sum_k {\cfrac{1}{\mathit{{\sigma}}_k^2}\cfrac{\partial\phi_c^k}{\partial R_\star}\cfrac{\partial\phi_c^k}{\partial{s}}}&
\sum_k {\cfrac{1}{\mathit{{\sigma}}_k^2}\cfrac{\partial\phi_c^k}{\partial s}\cfrac{\partial\phi_c^k}{\partial{r}}}&
\sum_k {\cfrac{1}{\mathit{{\sigma}}_k^2}\Bigl(\cfrac{\partial\phi_c^k}{\partial{s}}\Bigr)^2}
\end{bmatrix}
.
\end{equation}

We neglect now the detector noise (photon noise regime), and we assume 
that we are working outside the null 
regions, so that the phase closure may be approximated by
Eq.~\ref{clot}. Each element of the matrix above is the sum of nine
terms. There are two types of terms
\begin{eqnarray}
A_{ln,pq} & = & \sum_k f(u_{ln}^k){\rm sin}(2\pi u_{ln}^ks){\rm
  cos}(2\pi u_{pq}^ks) \\
B_{ln,pq} & = & \sum_k g(u_{ln}^k){\rm sin}(2\pi u_{ln}^ks){\rm
  sin}(2\pi u_{pq}^ks)
\end{eqnarray}
where $f(u)$ and $g(u)$ are smoothly varying functions over the
working frequency intervals $\Delta u_{ln}$. 

At last, we assume that the projected separation
$s$ is larger than a few stellar radii and that the working frequency
intervals are larger than $s^{-1}$, but smaller than the mean
frequencies:
\begin{equation}
s >{\rm a~few}~R_\star \qquad {\rm and}\qquad \bar{u}_{ln} \gg \Delta u_{ln} \ge s^{-1} .
\end{equation}
Under these conditions, all the $A$ terms are equal to zero and, given
that $u_{12} \neq u_{23} \neq u_{13}$ (non redundancy), the only non
null $B$ terms are those for which $\{ln\}=\{pq\}$. In addition,
\begin{equation}
B_{ln,ln} = \sum_k g(u_{ln}^k){\rm sin}^2(2\pi u_{ln}^ks) \approx
\cfrac{N_c}{2}\times\overline{g(u_{ln})}
\end{equation}
where $\overline{g(u_{ln})}$ is the average value of $g(u_{ln})$ over
the working frequency interval $\Delta u_{ln}$. The matrix
$^t\!\mathcal{M}\times\mathcal{M}$ may then be written into 
the form
\begin{eqnarray}
{}\!^t\mathcal{M}\times\mathcal{M} \approx
\frac{K}{N_{{\normalsubformula{\text{tel}}}}^2}
% \nonumber \\
\begin{bmatrix}
r^2\overline{X(R_\star)} & r\overline{Y(R_\star)} & 0 \\
 r\overline{Y(R_\star)}  & 1& 0 \\
 0 & 0 & (2\pi \bar{u}_{max} r)^2\overline{Z(R_\star)}
\end{bmatrix}
\end{eqnarray}
where $K$ is the total number of photoevents detected during the whole
observation and $\bar{u}_{max}$ the average value of the highest
frequency. $X(R_\star)$, $Y(R_\star)$, and $Z(R_\star)$ are given by
\begin{eqnarray}
\overline{X(R_\star)} = &  \cfrac{{\sum_{ln} \cfrac{1}{V^4_\star(u_{ln})} \Biggl (\cfrac{\partial V_\star(u_{ln})}{\partial R_\star} \Biggr )^2}}
%\Biggl/ 
{ \sum_{ln} \cfrac{1}{V^2_\star(u_{ln})} } \\
\overline{Y(R_\star)} = &  \cfrac{\sum_{ln} \cfrac{1}{V^3_\star(u_{ln})} \cfrac{\partial V_\star(u_{ln})}{\partial R_\star}}
%\Biggl/ 
{ \sum_{ln} \cfrac{1}{V^2_\star(u_{ln})} } \\
\overline{Z(R_\star)} = & \cfrac{ {\cfrac{1}{\bar{u}^2_{max}}\sum_{ln} \cfrac{u^2_{ln}}{V^2_\star(u_{ln})}}} 
%\Biggl/
{ \sum_{ln} \cfrac{1}{V^2_\star(u_{ln})}} 
.
\end{eqnarray}

The errors on the parameters of the system may now be
computed. Inverting the matrix $^t\mathcal{M}\times\mathcal{M}$
provides
\begin{eqnarray}
\sigma^2(R_\star) 
\approx & \cfrac{N_{{\normalsubformula{\text{tel}}}}^2}{r^2K} \times
\cfrac{1}{\overline{X(R_\star)}-\overline{Y(R_\star)}^2} \\
\sigma^2(r)
\approx & \cfrac{N_{{\normalsubformula{\text{tel}}}}^2}{K} \times
\cfrac{\overline{X(R_\star)}}{\overline{X(R_\star)}-\overline{Y(R_\star)}^2} \\
\sigma^2(s/R_\star)
\approx & \cfrac{N_{{\normalsubformula{\text{tel}}}}^2}{r^2K} \times 
\cfrac{1}{(2\pi \bar{u}_{max} R_\star)^2\overline{Z(R_\star)}} \, .
\end{eqnarray}

Let us focus on the flux ratio and on the stellar separation. Numerical
calculations show that the quantities
$\overline{X(R_\star)}/[\overline{X(R_\star)}-\overline{Y(R_\star)}^2]$
and $\overline{Z(R_\star)}$ are fairly constant and 
close to 1 in a wide range of experimental
conditions, even for observations performed around the nulls. Then,
in the photon noise regime, the
errors on the flux ratio and the separation may be fairly approximated
by
\begin{eqnarray}
\sigma(r) 
\approx & \cfrac{N_{{\normalsubformula{\text{tel}}}}}{\sqrt{K}} & \\
\sigma(s/R_\star) 
\approx & \cfrac{N_{{\normalsubformula{\text{tel}}}}}{r\sqrt{K}} & \times
\cfrac{1}{2\pi \bar{u}_{max} R_\star}  \, .
\end{eqnarray}
\begin{acknowledgements}
  This research has made use of the \texttt{SearchCal} service of the
  Jean-Marie Mariotti Centre\footnote{Available at http://jmmc.fr}, of
  the CDS
  Astronomical Databases SIMBAD and VIZIER, and of the NASA Astrophysics Data
  System Abstract Service.
\end{acknowledgements}

\bibliographystyle{aa} % style aa.bst
\bibliography{paperI} % your references Yourfile.bib

\end{document}